\documentclass[sigconf]{acmart}

\renewcommand\footnotetextcopyrightpermission[1]{}
\setcopyright{none}

\AtBeginDocument{\pagestyle{plain}}

\usepackage{booktabs}
\usepackage{multirow}
\usepackage{tabularx}
\usepackage{amsmath}

\usepackage{amssymb}
\usepackage{graphicx}
\setkeys{Gin}{draft=false}
\usepackage{subcaption}
\usepackage[normalem]{ulem}
\usepackage{xcolor}
\definecolor{darkgreen}{RGB}{0,120,0}



\begin{document}

\title{SCALE-Sim TPU: Validating and Extending SCALE-Sim for TPUs}

\author{Jingtian Dang}
\affiliation{
  \institution{Georgia Institute of Technology}
  \city{Atlanta}
  \state{Georgia}
  \country{USA}
}
\email{dangjingtian@gatech.edu}

\author{Ritik Raj}
\affiliation{
  \institution{Georgia Institute of Technology}
  \city{Atlanta}
  \state{Georgia}
  \country{USA}
}

\author{Changhai Man}
\affiliation{
  \institution{Georgia Institute of Technology}
  \city{Atlanta}
  \state{Georgia}
  \country{USA}
}

\author{Jianming Tong}
\affiliation{
  \institution{Georgia Institute of Technology}
  \city{Atlanta}
  \state{Georgia}
  \country{USA}
}

\author{Tushar Krishna}
\affiliation{
  \institution{Georgia Institute of Technology}
  \city{Atlanta}
  \state{Georgia}
  \country{USA}
}

\begin{abstract}
Cycle-accurate simulators are widely used to study systolic accelerators, yet their accuracy and usability are often limited by weak validation against real hardware and poor integration with modern ML compiler stacks. This paper presents \textsc{SCALE-Sim TPU}, a validated and extended version of SCALE-Sim v3 for TPU-style accelerators.
Specifically, we make three contributions:
(1) We validate SCALE-Sim’s systolic GEMM model against measurements on Google TPU v4 and show that simulated cycle counts exhibit a strong linear correlation with hardware latency, enabling a simple cycle-to-latency mapping.
(2) We introduce lightweight learned latency models for non-systolic elementwise operations, achieving median relative errors below 3\% using only tensor size and shape, substantially improving end-to-end latency estimation.
(3) We integrate a StableHLO-based frontend that allows workloads from modern ML frameworks such as JAX and PyTorch to be simulated directly via a unified compiler IR.
Together, these contributions improve the fidelity, coverage, and practicality of cycle-accurate simulation for whole-model performance analysis on TPUs.
\end{abstract}

\maketitle
\thispagestyle{plain}
\begin{figure}[t]
  \centering
  \includegraphics[width=\linewidth]{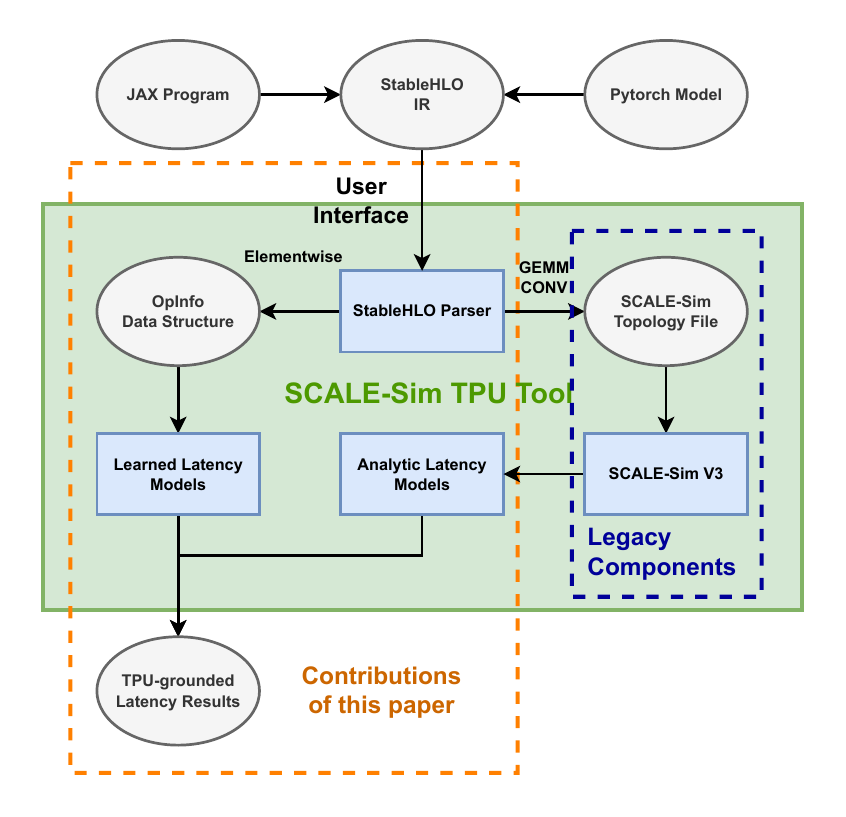}
  \caption{\textbf{Overview of the SCALE-Sim TPU Workflow and Contributions.} ML programs written in JAX or PyTorch are compiled into StableHLO, which serves as a unified input interface. The StableHLO parser extracts operator metadata and routes systolic-array operations (e.g., GEMM and convolution) to SCALE-Sim v3, while non-systolic operations are handled by lightweight analytic or learned latency models. Circles denote data artifacts or files, and rectangles denote processing components. The green region represents the complete SCALE-Sim TPU toolchain. The blue dashed region highlights legacy components inherited from prior work (SCALE-Sim v3), while the orange dashed region highlights the components introduced or extended in this paper.}
  \label{fig:overview}
\end{figure}
\section{Introduction}

Systolic-array accelerators such as Google’s Tensor Processing Units (TPUs) are a dominant substrate for modern machine learning workloads. Cycle-accurate simulators are therefore widely used to analyze performance bottlenecks, guide architectural exploration, and support benchmarking. However, in practice, two key limitations reduce the effectiveness of existing systolic-array simulators. First, their predictive validity is often unclear due to limited validation against real, production-scale hardware. Second, most simulators focus primarily on systolic kernels (e.g., GEMM and convolution), while ignoring or oversimplifying non-systolic operations, and require bespoke operator descriptions that are poorly aligned with modern ML compilation pipelines.

This paper addresses these gaps by validating and extending SCALE-Sim v3\cite{scalesimv3}, a modular cycle-accurate simulator for systolic accelerators. Our goal is to enable accurate, whole-model latency estimation for TPU-style accelerators directly from modern compiler IR, while maintaining lightweight modeling complexity.
Our contributions are:

\noindent $\bullet$ \textbf{TPU-grounded validation of SCALE-Sim's systolic array model.} We perform a TPU-grounded validation of SCALE-Sim’s systolic-array performance model by comparing simulated cycle counts against measured kernel latencies on Google TPU v4 across a wide range of GEMM shapes. We show that a simple analytical cycle model, when combined with a linear cycle-to-time mapping, can effectively translate SCALE-Sim cycle counts into TPU execution time, while also highlighting regimes where this approximation becomes less accurate.

\noindent $\bullet$ \textbf{Learnable latency modeling for elementwise operations.} To better account for the contribution of non-systolic operators to end-to-end model latency, we extend SCALE-Sim with lightweight, learnable latency models for elementwise operations. Using histogram-based gradient boosting regression, we model elementwise operator latency as a function of tensor size and shape. The resulting models achieve high predictive accuracy on TPU hardware, with median relative errors below 3\% for representative operators such as addition and ReLU, substantially improving whole-model latency estimation beyond systolic kernels alone.

\noindent $\bullet$ \textbf{StableHLO-based, framework-agnostic user interface for accelerator simulation.}
We introduce a StableHLO-based frontend that enables framework-agnostic simulation from modern ML stacks, including JAX and PyTorch. The frontend parses StableHLO to extract operator metadata and identifies operations supported by SCALE-Sim TPU, eliminating ad-hoc configuration formats and decoupling the simulator from framework-specific IRs. Supported systolic operations are routed to validated analytical models, while supported non-systolic operations are handled by learned latency models, enabling end-to-end latency estimation for the covered subset of compiler-emitted programs.

Together, these contributions improve the \textbf{fidelity}, \textbf{coverage}, and \textbf{usability} of cycle-accurate systolic-array simulation. SCALE-Sim TPU provides a practical toolchain for evaluating real ML workloads on TPUs, grounded in hardware measurements and compatible with modern compiler infrastructures.

\noindent Figure~\ref{fig:overview} summarizes the end-to-end workflow and highlights the components introduced in this paper.

\section{Background and Motivation}
\subsection{SCALE-Sim V3}
SCALE-Sim v3\cite{scalesimv3} is a modular, cycle-accurate simulator designed for performance analysis of systolic-array–based accelerators. It extends earlier versions of SCALE-Sim by supporting multi-core architectures with spatio-temporal partitioning, sparse matrix multiplication, detailed DRAM modeling via Ramulator, on-chip data layout effects, and energy estimation through Accelergy. These enhancements allow SCALE-Sim v3 to capture interactions between compute, memory, and data movement more faithfully, enabling detailed exploration of architectural trade-offs across latency, bandwidth, and energy for modern AI workloads.
\subsection{TPU v4}
TPU v4\cite{tpuv4} is Google’s fourth-generation Tensor Processing Unit, designed to accelerate large-scale machine learning workloads using systolic-array–based matrix multiplication. It features significantly increased compute density, higher-bandwidth on-chip and off-chip memory, and improved interconnects compared to earlier TPU generations, enabling efficient execution of large GEMM- and convolution-dominated workloads. TPU v4 serves as a production-scale accelerator platform and provides a representative target for validating cycle-level performance models against real hardware behavior.
\subsection{Motivation}
\paragraph{Why validate against real hardware?}
While simulator-based studies are common, many systolic-array simulators are rarely validated against production accelerators.
To ground SCALE-Sim's modeling assumptions, we use state-of-the-art systolic-array accelerator Google TPU as a representative platform and analyze when SCALE-Sim correctly reflects performance behavior.

\paragraph{Why model elementwise operations?}
End-to-end ML model latency is not determined solely by systolic-array kernels such as GEMM and convolution. Modern ML workloads contain a large number of non-GEMM operators, including elementwise additions, activations, normalization, and other memory-bound operations, whose cumulative latency can be significant. Recent workload characterization shows that non-GEMM operations can account for \textbf{11.3\%--73.6\%} of total inference time across models and platforms, and still contribute \textbf{15\%--48\%} of end-to-end latency even after operator fusion~\cite{karami_understanding_2025}. As GEMM and convolution are increasingly accelerated by specialized hardware, these non-systolic operators become a dominant performance bottleneck. Prior work on inference latency prediction therefore emphasizes the need for operation-level modeling, including elementwise operators, to achieve accurate end-to-end latency estimation across diverse hardware and software stacks~\cite{li_inference_2024}. Consequently, to enable faithful whole-model latency estimation, it is necessary to model non-systolic elementwise operations alongside GEMM and convolution.

\paragraph{Why StableHLO?}
Modern ML workloads are authored in frameworks such as JAX and PyTorch and increasingly flow through compiler IRs.
StableHLO provides a convenient, unified representation for extracting operator shapes and attributes, avoiding per-operator hand-written configuration formats and simplifying integration \cite{stablehlo2024}.

\section{Related Works}
Prior work on accelerator performance modeling has largely focused on either cycle-level simulation or analytical modeling of systolic architectures.
TimeLoop \cite{timeloop} offers a flexible analytical approach for mapping DNN layers to accelerator architectures; however, it relies on custom configuration files and validates results analytically (or against lower-level models) rather than through direct measurements on production hardware.

COCOSSim \cite{cocossim} performs cycle-accurate simulation of both systolic arrays and vector units, and reports comparisons against TPU v3 for selected kernels and neural-network layers.
Its vector-unit abstraction represents execution as sequences of reduction and broadcast phases, enabling accurate modeling of operators such as Softmax and Attention.
However, this phase-based abstraction is specialized and does not expose general elementwise operations as first-class kernels.

\begin{table*}[t]
  \centering
  \small
  \setlength{\tabcolsep}{4pt}
  \renewcommand{\arraystretch}{1.15}
  \caption{\textbf{Comparison of accelerator performance simulators and modeling frameworks.} We compare prior work along three axes: real hardware validation, support for elementwise (non-systolic) operations, and user interface. SCALE-Sim TPU (this work) is validated against TPU v4 hardware, extends SCALE-Sim v3 with explicit modeling of elementwise operators, and uses compiler-level StableHLO as a unified metadata interface, avoiding framework-specific input formats.}
  \label{tab:simulator-comparison}
  \begin{tabularx}{0.9\textwidth}{@{\extracolsep{\fill}} c c c c @{}}
    \toprule
    \textbf{Work} & \textbf{Real Hardware Validation} & \textbf{Elementwise Operations} & \textbf{User Interface} \\
    \midrule
    SCALE-Sim v3 \cite{scalesimv3} & \textcolor{red}{ No} & \textcolor{red}{No} & CSV \\

    TimeLoop \cite{timeloop} & \textcolor{red}{No}  & \textcolor{red}{No} & YAML \\

    COCOSSim \cite{cocossim} & \textcolor{darkgreen}{Yes(TPU v3)}  & \textcolor{red}{No}  & PyTorch \\

    \textbf{SCALE-Sim TPU (this work)}  & \textcolor{darkgreen}{\textbf{Yes (TPU v4)}} & \textcolor{darkgreen}{\textbf{Yes}} & \textbf{StableHLO} \\
    \bottomrule
  \end{tabularx}
\end{table*}

SCALE-Sim TPU builds on top of SCALE-Sim v3 and targets the capabilities missing from prior work.
By validating against TPU hardware, extending support beyond systolic kernels, and adopting StableHLO as the user interface, it enables kernel-level latency estimation directly from compiler IR.
Because StableHLO is widely used in MLIR-based compilation stacks, this interface provides a practical path to evaluating real workloads without manual topology specification.

\section{Methodology}
We organize the methodology around the three extensions in SCALE-Sim V3: (i) TPU-grounded GEMM validation, (ii) learnable latency modeling for non-systolic operators, and (iii) a StableHLO-based frontend.

\subsection{GEMM Modeling on Google TPU}

\subsubsection{Validation of SCALE-Sim Against TPU Latency}
\label{sec:gemm-validation}
We validate SCALE-Sim under configurations that are intentionally close to Google Cloud TPU v4 systolic-array execution models.\cite{tpuv4}

\paragraph{Setup and simulator configuration}
We configure SCALE-Sim to closely mirror TPU-style systolic execution: a 2D systolic array with a \emph{128\,$\times$\,128} MAC mesh (matching TPU v4).
We evaluate GEMM workloads of the form $\mathbf{C}=\mathbf{A}\mathbf{B}$ with $(M,K)\times(K,N)$ shapes using a structured parameter sweep.

For each regime, we sweep \emph{each of the three dimensions} $(M,K,N)$ over the same range separately with a fixed step size: 
\begin{itemize}
  \item \textbf{Small:} 32--128 with step size 16
  \item \textbf{Medium:} 128--1024 with step size 128
  \item \textbf{Large:} 1024--4096 with step size 512
\end{itemize}
The three size regimes are chosen to reflect distinct execution behaviors of TPU-style systolic arrays without implying capacity overflow of on-chip storage.
The \emph{small} regime (32--128) captures cases where matrix dimensions are comparable to or smaller than the array size, leading to under-utilization and making pipeline fill and drain effects dominant.
The \emph{medium} regime (128--1024) spans shapes that moderately exceed the array dimensions, where utilization improves and steady-state systolic execution becomes the primary performance determinant.
The \emph{large} regime (1024--4096) represents workloads that, while still fitting within on-chip storage constraints, require increasingly complex tiling, accumulation, and scheduling decisions by the TPU compiler.
In this regime, execution behavior is shaped less by raw array occupancy and more by how computation is partitioned and overlapped across tiles. 

Smaller step sizes are used in the small regime to capture fine-grained utilization effects, whereas coarser steps suffice in the medium and large regimes where performance trends vary more smoothly with scale.

\paragraph{Collecting SCALE-Sim cycles and TPU GEMM kernel runtimes.}
For each GEMM shape, we record (1) the predicted cycle count from SCALE-Sim and (2) the measured kernel runtime on TPU v4.
The measured runtime reflects on-chip execution only, excluding HBM-to-core data transfers, thereby isolating the core compute behavior of the TPU’s systolic array.

\paragraph{Linear regression for validation (cycle-to-time calibration).}
We fit a simple linear model that maps SCALE-Sim cycles to TPU fusion time:
$\hat{t} = \alpha \cdot \texttt{cycles} + \beta$,
where $\alpha$ represents the effective time per simulated cycle, $\alpha \cdot \texttt{cycles}$ captures the cycle-dependent execution time, and $\beta$ captures fixed overheads not modeled by SCALE-Sim.
We report regression fits separately for the three size regimes to highlight regime-dependent behavior.

\begin{figure*}[t]
  \centering

  \begin{subfigure}[t]{0.32\textwidth}
    \centering
    \includegraphics[width=\linewidth]{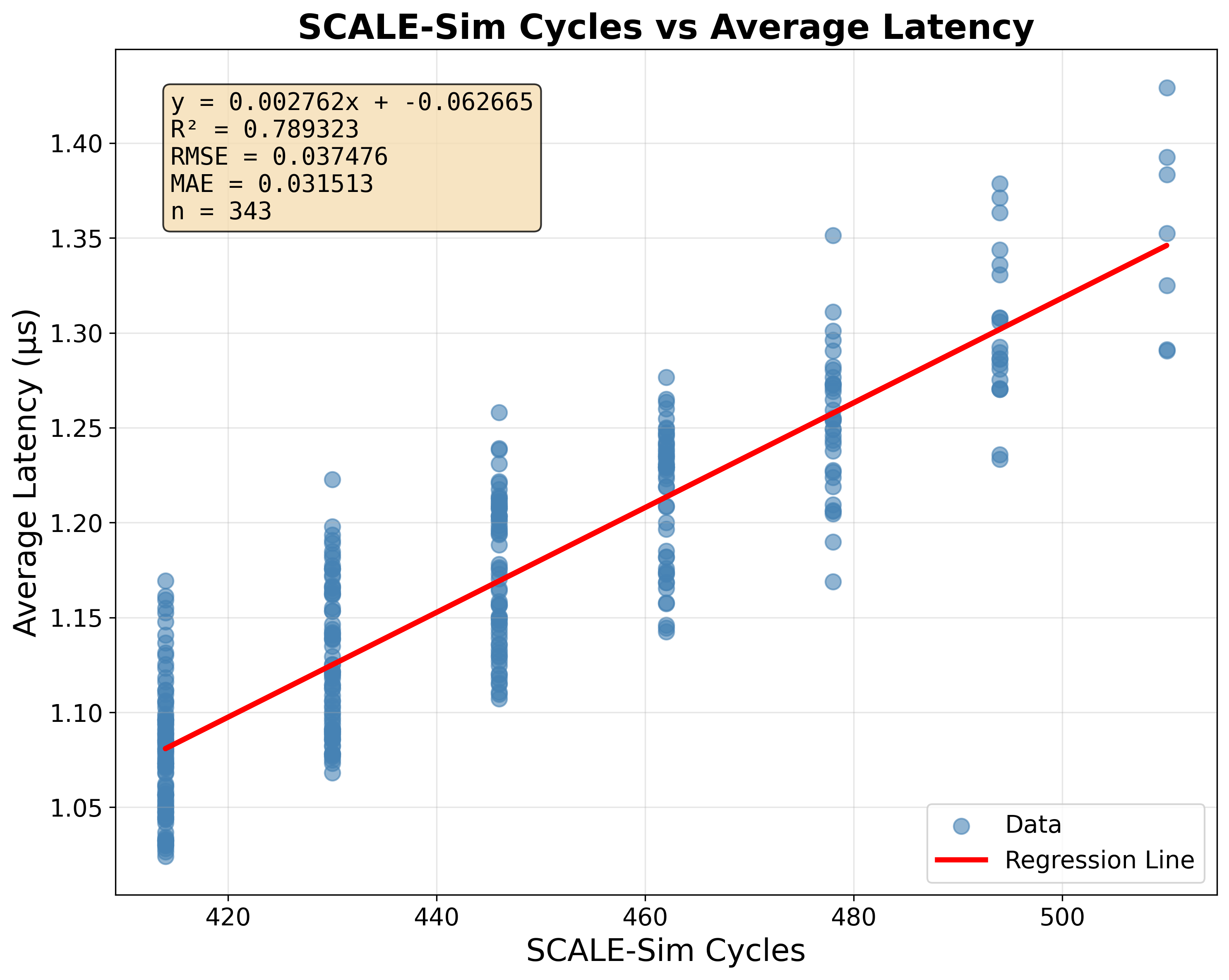}
    \caption{Matrix Size in Small regime}
  \end{subfigure}\hfill
  \begin{subfigure}[t]{0.32\textwidth}
    \centering
    \includegraphics[width=\linewidth]{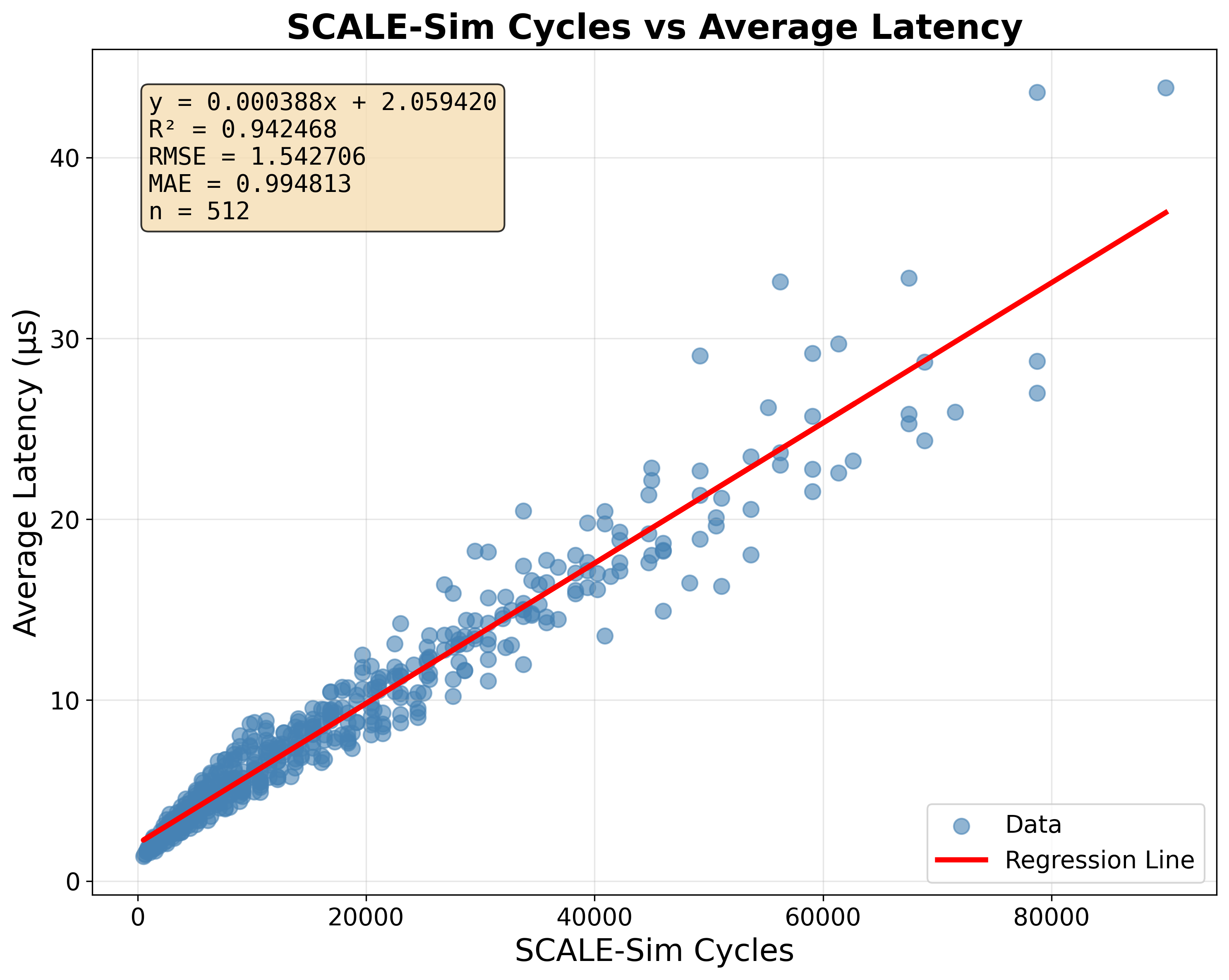}
    \caption{Matrix Size in Medium regime}
  \end{subfigure}\hfill
  \begin{subfigure}[t]{0.32\textwidth}
    \centering
    \includegraphics[width=\linewidth]{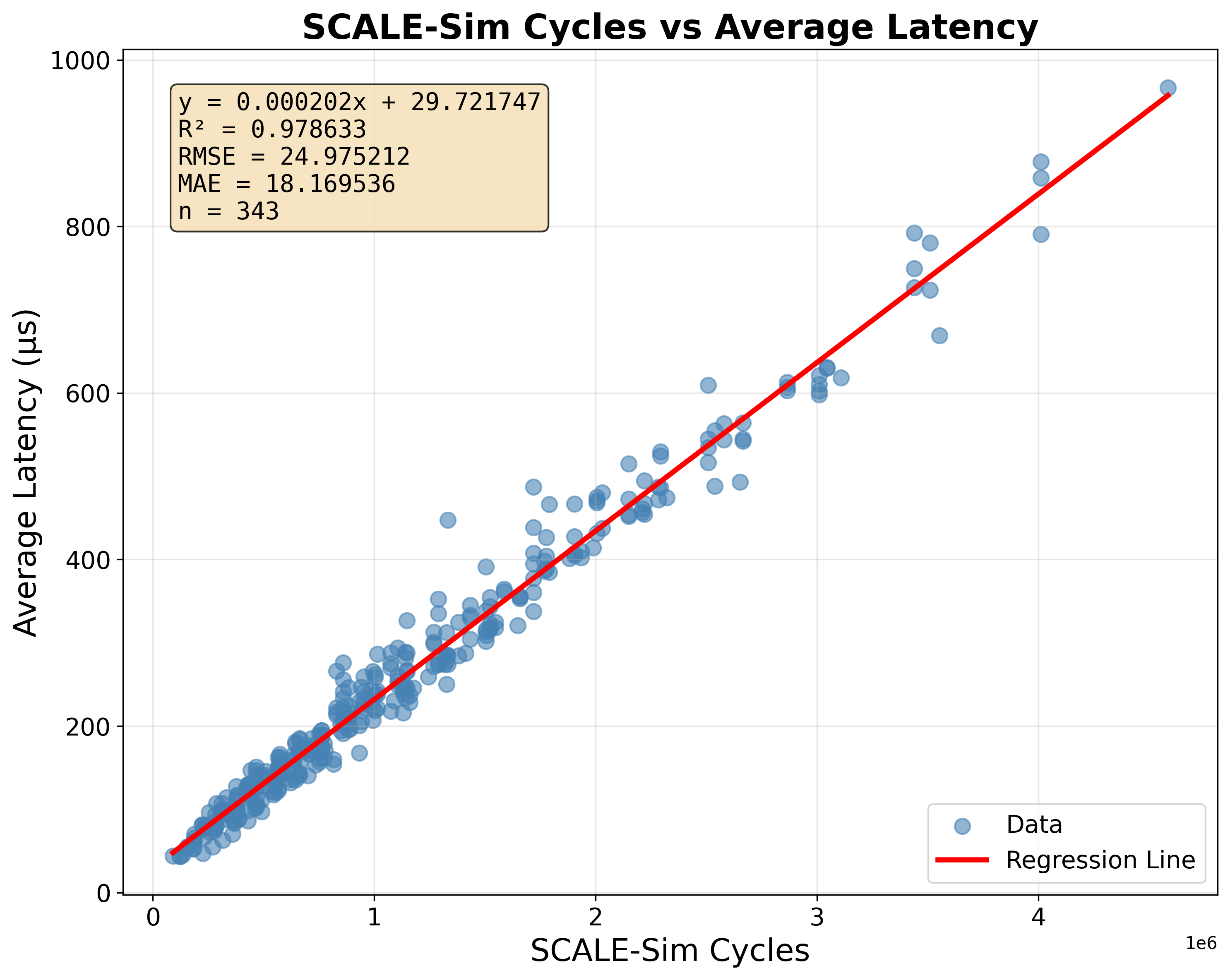}
    \caption{Matrix Size in Large regime}
  \end{subfigure}

  \caption{SCALE-Sim--to--TPU v4 regression for systolic GEMM across three size regimes.
    Each plot shows SCALE-Sim predicted cycle counts (\emph{x}-axis) versus measured TPU v4 kernel latency (\emph{y}-axis) for GEMM workloads executed on a $128\times128$ systolic array.
    Each point corresponds to one GEMM shape from the sweep, and the solid line shows a least-squares linear regression; the inset reports $R^2$, RMSE, MAE, and sample count.
    Across all regimes, SCALE-Sim cycle counts show a clear linear relationship with measured TPU execution time, indicating that simulated cycles provide a useful predictor of systolic GEMM latency.
  }
  \label{fig:v4-reg}
\end{figure*}

\paragraph{Linear regression results.}
As shown in Fig.~\ref{fig:v4-reg}, SCALE-Sim cycle predictions exhibit a consistent linear relationship with measured TPU v4 fusion-kernel latency across all three size regimes.
The fitted regressions achieve moderate to high goodness-of-fit, with $R^2$ values ranging from approximately 0.79 in the small regime to over 0.97 in the medium and large regimes.
Here, $R^2$ measures the fraction of variance in the measured TPU latency explained by the linear model, with higher values indicating that simulated cycle counts account for a larger portion of the observed runtime variation.
These results suggest that simulated cycle counts capture a substantial component of the execution cost of TPU systolic GEMM across a wide range of problem sizes.


\paragraph{Error analysis and insights}
Prediction errors are smallest in the medium regime, where array utilization is high and execution behavior closely matches the idealized systolic model.
In the small regime, under-utilization and pipeline effects introduce greater relative variability, leading to a lower $R^2$ despite small absolute errors.
In the large regime, increased variance arises from factors outside SCALE-Sim's compute-centric model, including TPU compiler tiling decisions, layout transformations, and limits on memory bandwidth and on-chip buffering.

Overall, the observed errors are well-behaved and scale smoothly with problem size, indicating no strong systematic bias across the evaluated GEMM shapes.

\subsubsection{Enabling Direct TPU Latency Estimation}
Based on the strong and consistent linear correlation observed across all size regimes (Section~\ref{sec:gemm-validation}), we extend SCALE-Sim to directly report estimated TPU execution time for systolic GEMM and convolution kernels.
Specifically, we reuse the regime-specific linear regression functions obtained during validation to map simulated cycle counts to wall-clock latency.
Specifically, we reuse the linear regression function obtained during validation to map simulated cycle counts to wall-clock latency.

This extension does not introduce additional modeling assumptions or parameters beyond those already used for validation.
Instead, it leverages the observed relationship to provide users with latency estimates in physical time units, eliminating the need for manual post-processing or external calibration.
The regression parameters are specific to the target TPU platform and can be re-derived when targeting different TPU generations.

With this extension, SCALE-Sim TPU outputs both cycle-level and time-based performance estimates, enabling practical kernel-level latency analysis and comparison directly from StableHLO-derived workloads.

\subsection{Learnable latency model for elementwise operations}


\paragraph{Target operators}
We target common non-systolic operators that are pervasive in end-to-end ML workloads but are not executed on the TPU's systolic array, with an initial focus on elementwise operations (e.g., add, multiply, ReLU).

\begin{figure}[t]
  \centering

    \begin{subfigure}[t]{\linewidth}
    \centering
    \includegraphics[width=\linewidth]{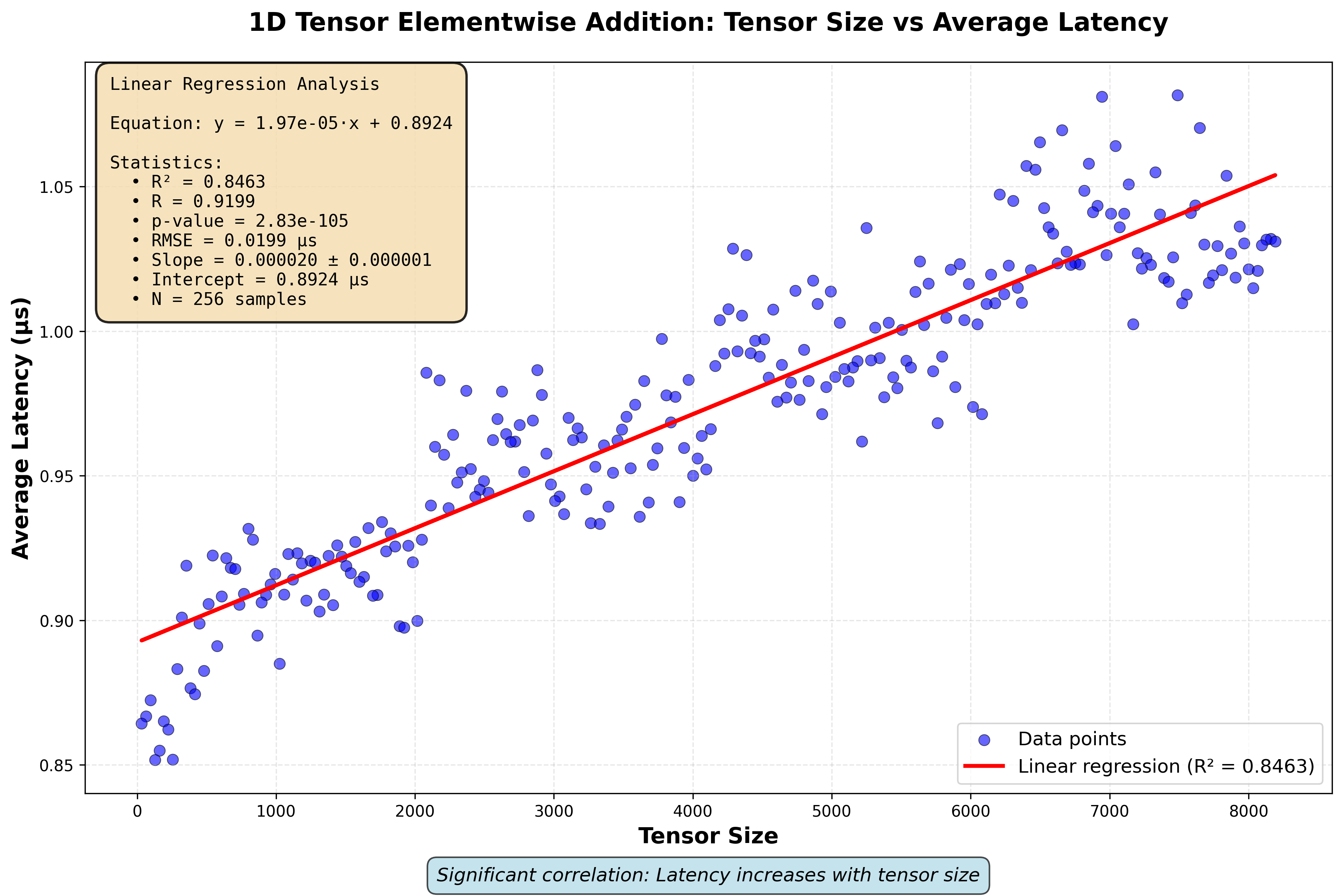}
    \caption{1D sweep.}
    \label{fig:elemwise-1d}
  \end{subfigure}
  
  \vspace{0.5em}

  \begin{subfigure}[t]{\linewidth}
    \centering
    \includegraphics[width=\linewidth]{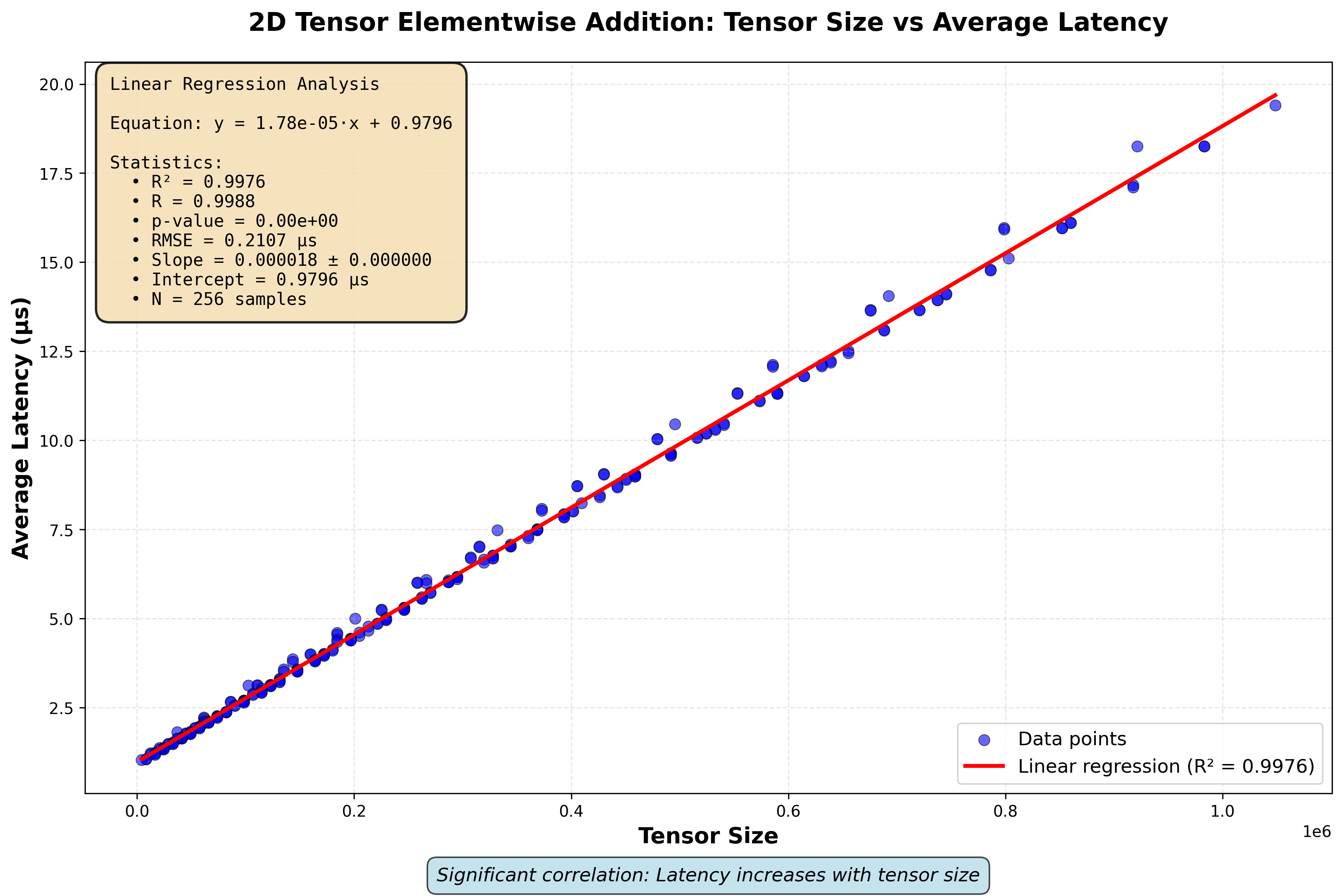}
    \caption{2D sweep.}
    \label{fig:elemwise-2d}
  \end{subfigure}

  \caption{bf16 elementwise-add latency vs. tensor size for 1D (32--8192 step 32) and 2D (64--1024 step 64 per dim) sweeps; near-linear scaling with minor shape-dependent fluctuations.}
  \label{fig:elemwise-add}
\end{figure}

\paragraph{Exploratory analysis and motivation}
As an initial exploratory step, we manually collect bf16 elementwise-addition measurements over controlled 1D and 2D tensor-shape sweeps to understand how latency scales with tensor size.
For 1D tensors, we sweep length from 32 to 8192 (step 32).
For 2D tensors, we sweep each dimension from 64 to 1024 (step 64).
Figures~\ref{fig:elemwise-1d} and~\ref{fig:elemwise-2d} plot average latency versus tensor size for the 1D and 2D sweeps. 

Across both sweeps, latency is approximately linear in tensor size, indicating that runtime is largely driven by the number of processed elements.
However, we also observe small fluctuations: different shapes with the same tensor size can yield slightly different latencies.
This motivates including tensor-shape information (in addition to total element count) in our learnable non-systolic latency model. 

\paragraph{Feature selection}
We use tensor size and tensor shape as input features.
Tensor size captures the dominant scaling behavior for elementwise operations, since each element contributes a fixed amount of computation and memory traffic.
However, the measurements reveal consistent latency differences across shapes with the same size, implying that performance is also influenced by shape-related factors such as vectorization granularity, alignment, and scheduling thresholds.
Including tensor shape enables the model to learn these structured deviations.
Both features are statically known (compile-time metadata) and independent of runtime values, making them suitable for lightweight simulator integration. 

\paragraph{Model choice}
We use a Histogram-based Gradient Boosting Regressor (HGBR) to predict elementwise operator latency.
Gradient-boosted decision trees are data-efficient and robust to noise: they iteratively fit simple decision trees to residual errors and aggregate them into a strong predictor~\cite{friedman_greedy_2001}.
By construction, tree-based models naturally capture non-linear, non-smooth, and piecewise behaviors, making them well-suited for modeling performance discontinuities induced by hardware boundaries such as tiling, alignment, and resource thresholds.
Histogram-based gradient boosting further improves efficiency by discretizing continuous features into bins, reducing computation and memory overhead while preserving predictive accuracy~\cite{ke_lightgbm_2017}.
Compared to a single linear model, HGBR can represent discrete transitions and piecewise trends, while avoiding the larger data requirements and tuning complexity typically associated with neural network models.

\paragraph{Training data}
We construct the training dataset by measuring elementwise operator latency across a diverse set of bf16 tensor shapes, with total tensor sizes sampled log-uniformly up to a fixed maximum (up to \textasciitilde16M elements).
To capture shape-dependent performance effects, the dataset includes multiple factorizations of the same total element count, enabling the model to distinguish between tensors with identical sizes but different dimensional layouts.

In addition, we intentionally include shapes near common hardware-relevant boundary conditions, such as dimensions aligned to powers of two ($2^n$).
These boundary cases can trigger different compiler scheduling decisions, vectorization strategies, and memory alignment behavior on TPU hardware, leading to measurable latency differences even for tensors with similar sizes.
By incorporating such cases, the dataset exposes the latency model to both smooth scaling trends and discontinuities introduced by alignment- and scheduling-related effects.

For each shape, latency is measured multiple times and we use the median to reduce run-to-run noise. 

\paragraph{Training and validation protocol}
We train the model on a subset of tensor sizes and evaluate it on previously unseen sizes to assess generalization. The model is optimized to minimize absolute latency error in real time units. We report both absolute and relative error on the validation set to capture accuracy across small and large tensors. 

\subsection{StableHLO interface}
\paragraph{Motivation and overview}
StableHLO is a standardized intermediate representation (IR) used by modern ML frameworks such as JAX and PyTorch to describe accelerator-executable programs.
By targeting StableHLO as the input interface, SCALE-Sim TPU enables evaluation of realistic ML workloads without manually translating programs into custom SCALE-Sim topology files, lowering the barrier to use and improving reproducibility across workloads and toolchains.

Figure~\ref{fig:overview} provides a high-level overview of the SCALE-Sim TPU workflow.
ML programs written in JAX or PyTorch are compiled into StableHLO, after which a StableHLO parser extracts operation-level metadata and converts it into simulator-compatible representations.
Systolic-array operations are mapped to SCALE-Sim v3 using analytic performance models, while non-systolic operations are handled by lightweight learned latency models.

\paragraph{StableHLO parsing}
The StableHLO parser processes compiler-emitted StableHLO modules and extracts a lightweight representation for each operation.
For every StableHLO operation, the parser records:
\begin{itemize}
  \item the operation type (e.g., \texttt{dot\_general}, \texttt{convolution}, \texttt{add});
  \item the input and output tensor shapes;
  \item the data types (e.g., bf16);
  \item and relevant attributes, such as convolution parameters.
\end{itemize}
This information is stored in a uniform internal data structure (\texttt{OpInfo}), which decouples the frontend IR from backend performance models and enables extensible operator handling.

\paragraph{Operation conversion}
Parsed StableHLO operations are classified and converted into simulator-level operators based on their execution characteristics.

\paragraph{Dot to GEMM}
StableHLO \texttt{dot\_general} operations that match matrix-multiplication patterns are converted into GEMM operators.
The input tensor shapes and contracting dimensions are used to derive the corresponding $M$, $N$, and $K$ parameters for $\mathbf{C}=\mathbf{A}\mathbf{B}$, which are then passed to SCALE-Sim's GEMM modeling interface.
This conversion enables direct reuse of SCALE-Sim's validated systolic-array performance model.

\paragraph{Convolution}
StableHLO \texttt{convolution} operations are mapped to SCALE-Sim convolution inputs when applicable.
Tensor shapes and convolution attributes (e.g., strides, filter dimensions, and channel counts) are translated into SCALE-Sim convolution parameters, including IFMAP dimensions, filter sizes, number of channels, and number of filters.
This mapping preserves the structural information required by SCALE-Sim's convolution model, while excluding preprocessing and data-layout transformation costs that are not explicitly modeled.

\paragraph{Other operations}
Operations that are not executed on systolic arrays are currently classified as elementwise operations, including addition, subtraction, multiplication, maximum, and minimum.
For these operations, the parser extracts tensor-shape information and routes them to learned latency models.
This approach enables efficient modeling of elementwise operations that are common in modern ML workloads but are poorly captured by purely analytic models.


\section{Evaluation}
We evaluate SCALE-Sim TPU along three axes aligned with our contributions.
GEMM validation results (including linear regression diagnostics) are presented in Section~\ref{sec:gemm-validation}.

\subsection{Cycle-to-latency mapping evaluation (GEMM)}
\label{sec:cycle-to-latency-eval}

\begin{figure}[t]
  \centering
  \IfFileExists{Figures/validation/new_predicted_vs_actual.png}{
    \includegraphics[width=\linewidth]{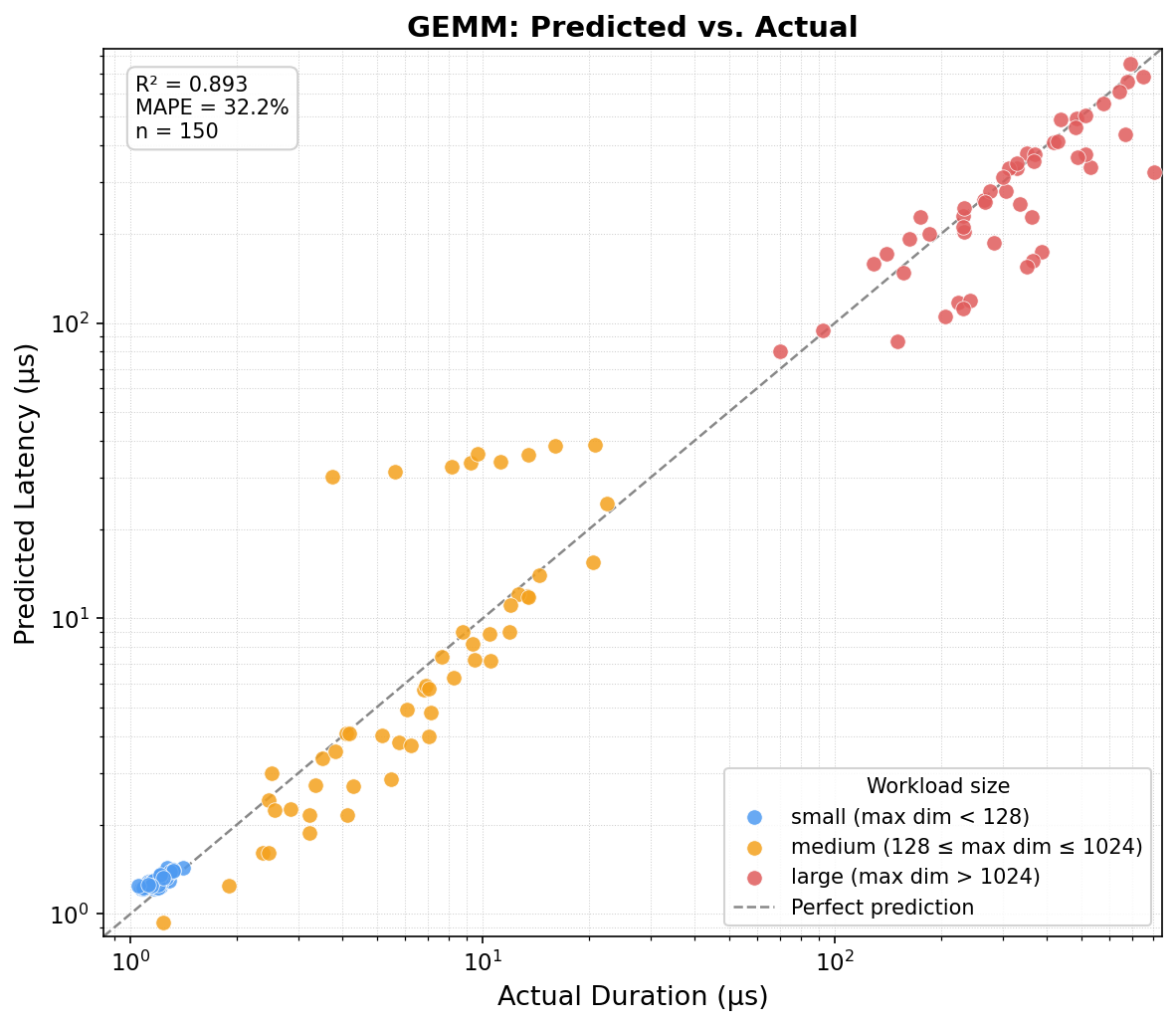}
  }{
    \fbox{\parbox{0.95\linewidth}{\centering Missing figure file}}
  }
  \caption{\textbf{Predicted vs. actual GEMM latency on TPU v4.} Each point represents a GEMM configuration, grouped by workload size (small, medium, large). The dashed line indicates perfect prediction ($y=x$). While SCALE-Sim TPU preserves the overall scaling trend across sizes ($R^2=0.893$), deviations are more pronounced for medium-sized workloads, leading to higher aggregate error (MAPE = 32.2\%).}
  \label{fig:cycle-to-latency}
\end{figure}

We evaluate the accuracy of the cycle-to-latency mapping for GEMM kernels by comparing predicted latency against measured TPU execution time across a range of matrix sizes. As shown in Figure~\ref{fig:cycle-to-latency}, the predictions exhibit a clear positive correlation with measured latency ($R^2=0.893$), indicating that SCALE-Sim captures the dominant scaling behavior of systolic-array execution. For small and large workloads, most points lie close to the ideal $y=x$ line, suggesting good agreement in these regimes. However, larger deviations are observed for medium-sized workloads, where predicted latency tends to underestimate or overestimate runtime. These discrepancies likely arise from unmodeled system-level effects, such as scheduling overheads, data movement, and kernel fusion behavior, which are not captured by the simple linear mapping. As a result, while the mapping provides a reasonable first-order estimate of TPU latency, its accuracy varies across workload regimes, and aggregate metrics such as MAPE (32.2\%) are influenced by these mid-range deviations.

\subsection{Elementwise latency prediction}
\label{sec:nonsystolic-eval}

\paragraph{Evaluation of learned latency models for elementwise operations}
\begin{figure}[t]
  \centering

  \begin{subfigure}[t]{0.95\linewidth}
    \centering
    \includegraphics[width=\linewidth]{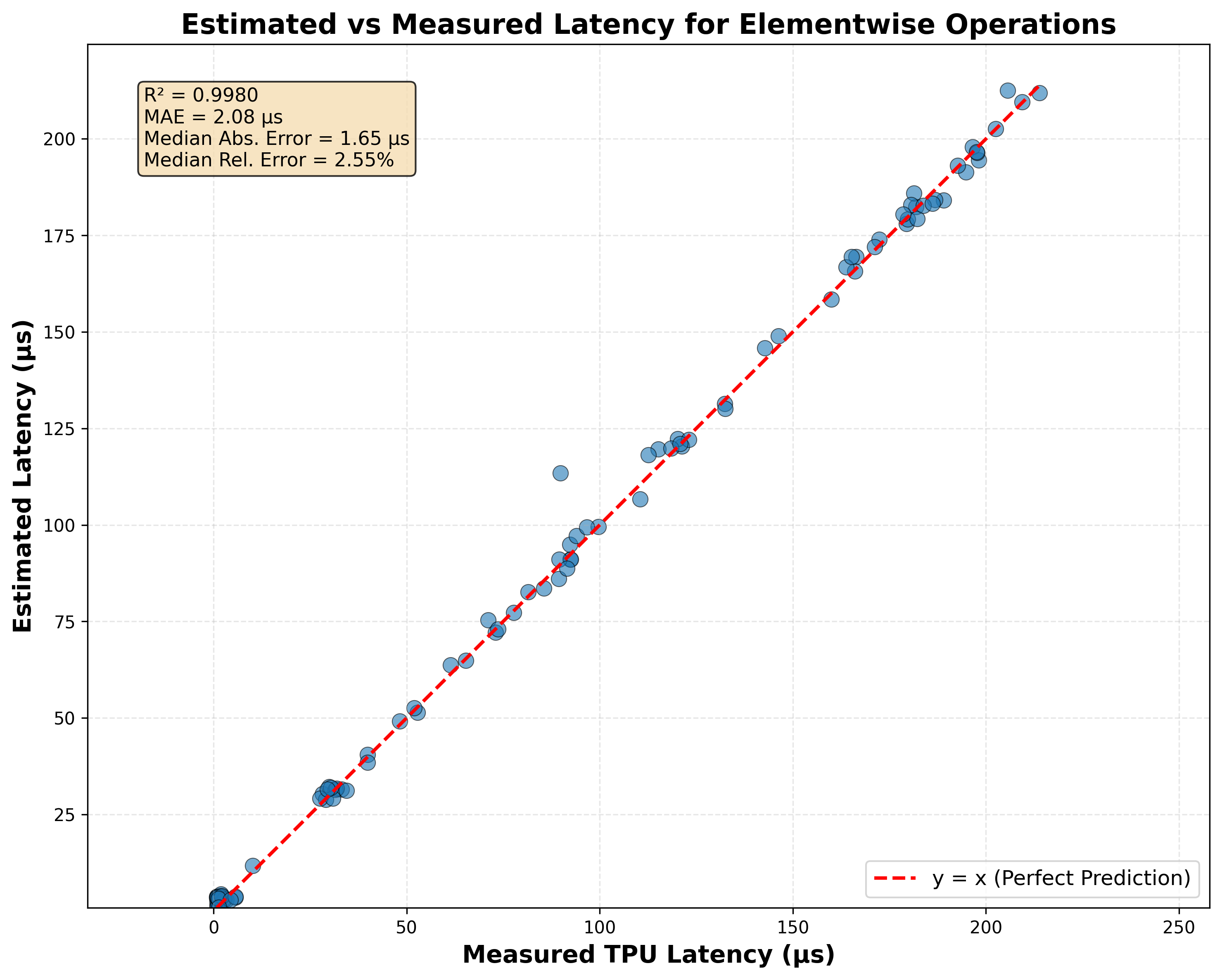}
    \caption{Elementwise addition.}
    \label{fig:model-eval-add}
  \end{subfigure}

  \par\medskip

  \begin{subfigure}[t]{0.95\linewidth}
    \centering
    \includegraphics[width=\linewidth]{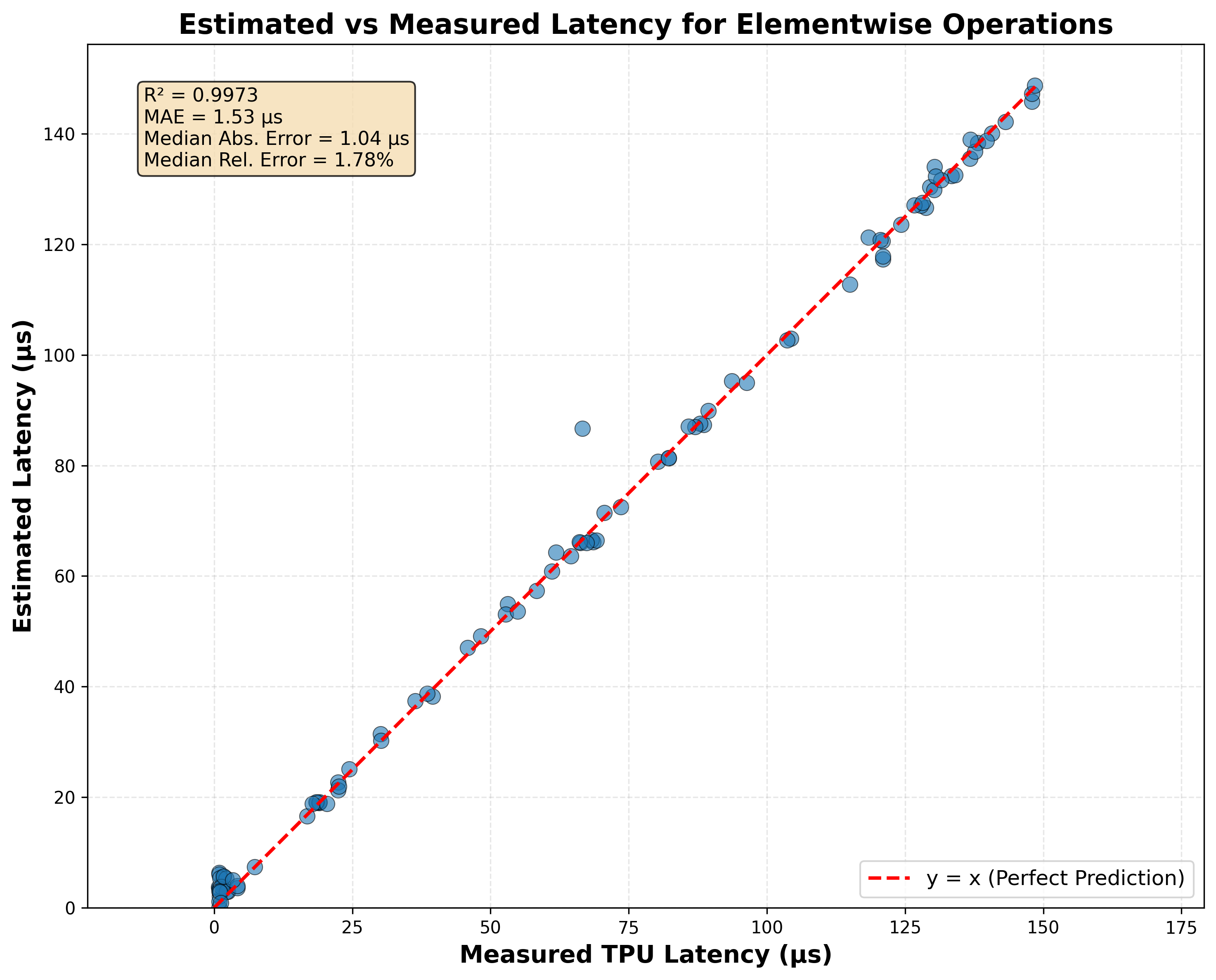}
    \caption{ReLU (maximum).}
    \label{fig:model-eval-relu}
  \end{subfigure}

  \caption{\textbf{Learned latency model evaluation for non-systolic (elementwise) operations.} Estimated versus measured TPU latency for (top) elementwise addition and (bottom) ReLU (maximum) across a diverse set of tensor shapes. Each point is one tensor shape; the dashed diagonal indicates perfect prediction.}
  \label{fig:model-eval}
\end{figure}

We evaluate the learned latency models for non-systolic operations using representative elementwise kernels, specifically addition and ReLU (maximum).
These operations capture two dominant classes of elementwise behavior: pure arithmetic and arithmetic with comparison.
For each operation, we compare the estimated latency produced by the learned model against measured TPU latency across a diverse set of tensor shapes.

Figure~\ref{fig:model-eval} shows the estimated versus measured latency for elementwise addition and ReLU.
Each point corresponds to one tensor shape, and the dashed diagonal indicates perfect prediction.
In both cases, the predicted latency closely tracks the measured latency across the full range of tensor sizes.

For elementwise addition, the learned model achieves an $R^2$ value of 0.9973, with a median absolute error of 1.04~$\mu$s and a median relative error of 1.78\%.
For ReLU, which introduces additional comparison and control logic, the model attains an $R^2$ of 0.9980, with a median absolute error of 1.65~$\mu$s and a median relative error of 2.55\%.
These results indicate that the model captures both the dominant size-dependent scaling behavior and secondary shape-related effects.

Across both operations, larger absolute errors occur primarily at small tensor sizes, where fixed overheads dominate execution time.
Nevertheless, relative error remains low across the evaluated range, demonstrating that the learned model generalizes well beyond the specific shapes used during training.




\section{Conclusion}
This paper validates and extends SCALE-Sim v3 for TPU-oriented performance analysis. By comparing against real TPU measurements, we show that SCALE-Sim’s cycle-level GEMM predictions exhibit a strong linear relationship with hardware execution time, enabling a simple and effective cycle-to-latency mapping. Building on this observation, we extend SCALE-Sim to directly report TPU latency estimates and to model common non-systolic elementwise operations using lightweight learned models. We further integrate StableHLO as a frontend, allowing workloads from modern ML compiler stacks to be simulated without manual topology specification. Together, these extensions improve the accuracy and usability of SCALE-Sim for whole-model latency estimation on TPUs.


\bibliographystyle{ACM-Reference-Format}
\bibliography{refs}

\end{document}